\documentclass[12pt,a4paper,english,superscriptaddress,aps,nofootinbib]{revtex4}
\usepackage{amsmath,amssymb,graphicx}
\makeatletter
\usepackage{babel}
\usepackage[active]{srcltx}
\usepackage{graphicx,color}
\usepackage{changebar}
\usepackage{hyperref}
\hypersetup{
	colorlinks=true,
	urlcolor= blue,
	citecolor=blue,
	linkcolor= blue,
	bookmarks=true,
	bookmarksopen=false,}

\usepackage[T1]{fontenc}
\usepackage{ae}
\usepackage[ansinew]{inputenc}
\usepackage{amsmath}
\usepackage{braket}
\usepackage{mathtools}
\usepackage{slashed}
\usepackage{empheq}
\usepackage{tikz}
\usepackage{multirow}
\usetikzlibrary{decorations.pathmorphing}
\usepackage[caption = false]{subfig}
\usepackage{graphicx}
\usepackage{amsfonts}
\usepackage{amsmath}
\newcommand{\mat}[1]{\mbox{\boldmath{$#1$}}}

\newcommand{\be}{\begin{equation}}
	\newcommand{\ee}{\end{equation}}
\newcommand{\bea}{\begin{eqnarray}}
	\newcommand{\eea}{\end{eqnarray}}

\begin{document}
	\title{Differential configurational complexity and phase transitions of the BPS solutions in the O(3)-sigma model}
	
\author{F. C. E. Lima}
\email[]{E-mail: cleiton.estevao@fisica.ufc.br}
\affiliation{Universidade Federal do Cear\'{a} (UFC), Departamento do F\'{i}sica - Campus do Pici, Fortaleza, CE, C. P. 6030, 60455-760, Brazil.}

\author{C. A. S. Almeida}
\email[]{E-mail: carlos@fisica.ufc.br}
\affiliation{Universidade Federal do Cear\'{a} (UFC), Departamento do F\'{i}sica - Campus do Pici, Fortaleza, CE, C. P. 6030, 60455-760, Brazil.}

\begin{abstract}
Using a spherically symmetric ansatz, we show that the Chern-Simons O(3)-sigma model with a logarithmic potential admits topological solutions. This result is quite interesting since the Gausson-type logarithmic potential only predicted topological solutions in $(1+1)$D models. To accomplish our goal, the Bogomol'nyi-Prasad-Sommerfield (BPS) method is used, to saturate the energy and obtain the BPS equations. Next, we show by the numerical method is the graphical results of the topological fields, as well as, the magnetic field behavior that generates a flux given by $\Phi_{flux}=-\mathcal{Q}/\kappa$ and the energy density of the structures of vortices. On the other hand, we evaluate the measure of the differential configurational complexity (DCC) of the topological structures, by considering the energy density of the vortex. This analysis is important because it will provide us with information about the possible phase transitions associated with the localized structures and it shows that our model only supports one phase transition.
\end{abstract}

\maketitle
\thispagestyle{empty}
\newpage
\section{Introduction}
\label{intro}
 The non-linear sigma model is a set of scalar fields mapped in a target space, where the fields are coordinates. In 1981, Alvarez-Gaumme and D. Freedman \cite{alvarez} used the target space as a Riemannian manifold for gauging of the space's isometries. Subsequently, other works consolidated the method of Alveraz-Gaumme and D. Freedman \cite{hull1,hull2,jack,CA}. The initial motivation of the research on the sigma model was to apply it to conformal field theory and superstring models. Posteriorly, Schroers \cite{schroers} investigates topological solitons using a breaking the scale invariance by gauging the subgroup U(1) of the O(3)-sigma model. In their paper, to build a field theory with BPS properties, the Maxwell term and scalar potential were used. Soon after, P. Ghosh and S. Ghosh \cite{ghosh} included a Chern-Simons term in the model. Several works in the following years treat the subject under many viewpoints \cite{muk1,CA2,rothe}. Recently,  the gauged O(3)-sigma model was readdressing in the context of Lorentz symmetry breaking models \cite{casana1,casana2} and in several approaches of new kinds of Bogomol'nyi-Prasad-Sommerfield (BPS) solutions \cite{casana2020,barto}.
 
It is interesting to mention here that the gauging of symmetry is used in the approach of the so-called Skyrme model to break the scale invariance. In principle, the Skyrme model is the non-linear sigma model generalized as the sigma model is invariant under the group SO(3), the Skyrme model will be too. In fact, for the Skyrme model to be invariant under the group SO(3), gauging the subgroup U(1) is necessary. This gauging will generate a coupling of the scalar field to the gauge field \cite{Gladi}. The result is a (3+1)D Skyrme-Maxwell model, which is a rich physical model predicting the proton-neutron mass difference \cite{durgut}. In (2+1)D, the version of the Skyrme model is known as the baby-Skyrme model\cite{piette}. 

The BPS approach \cite{Bogomol'nyi} allows us to study the field solutions topological and non-topological. Indeed, the BPS method is a series of inequalities of a homotopic class of solutions of a partial differential equation in the spatial infinite. In the context of topological defects, the BPS limit is represented by a saturation limit of the energy of the model \cite{Bogomol'nyi,Atmaja}. When the energy reaches the saturation limit, the equations of motion are known as BPS equations, and the order of these equations is reduced from second to first order \cite{Bogomol'nyi}.

Several works have considered the dynamics of the Chern-Simons gauge field to study the BPS solutions \cite{ghosh,Arthur,Gladi2,Kimm,Lima,Lima1}. These systems become self-dual for a suitable potential \cite{Jackiw,Jackiw1,LA11,LA12}, so their topological \cite{Horvathy} and non-topological solutions \cite{Spruck} will be present in the model.

In this work, we will introduce for the first time a logarithmic potential in the $O(3)$-sigma model context. The logarithmic potential produces solitonic solutions in bidimensional models \cite{Belendryasova}. Recently the studies indicate that the logarithmic potential can generate ring-like vortices with intense magnetic flux when the Higgs field lies coupled to the gauge field \cite{Lima2,LPA1,LPA2}. However, the first works on relativistic theories are due by G. Rosen \cite{Rosen,Rosen1}, Bialynicki-Birula, and Mycielski (BBM) in the context of Schr\"{o}dinger theories \cite{BBM}. For their Gaussian shape, BBM called Gaussons the soliton-like solutions of the Schr\"{o}dinger equation with logarithmic nonlinearity. Inspired by them, we called the Chern-Simons O(3)-sigma model with a logarithmic potential of Chern-Simons-Gausson O(3)-sigma model.

On the other hand, we can find some applications of the logarithmic potential in several models \cite{Zloshch1,Zloshch2,Zloshch3}. An attractive application of this theory arises when we consider an electrical charge with a self-consistent configuration used for describing logarithmic quantum Bose liquids \cite{Vladimir}. In this case, the electromagnetic field interacts with the physical vacuum. Also, models with the so-called gausson logarithmic term were used to describe theories of quantum gravity \cite{Konstantin1}, as well as quantum effects in nonlinear quantum theory \cite{Konstantin2}.

We introduced, throughout the work, the discussion of the differential configurational complexity (DCC) of BPS vortices in an Abelian model. DCC is a variant of configurational entropy (CE) that appeared in Ref. \cite{Gleiser3}, underpinned by the information theory of Claude E. Shannon \cite{Shannon}. From a quantum-mechanical viewpoint, Shannon's entropy is interested in giving us information on the probability of a particle evolving from one quantum state to another \cite{LRC}. CE and its variants are responsible for offering us information about the stability measure applied to a localized structure. The applications of this theory are vast. For example, the CE is used to investigate stable Q-ball solutions \cite{Gleiser4,Gleiser5} at the Chandrasekhar limit for white dwarfs, in the study of the non-equilibrium dynamics of spontaneous symmetry breaking \cite{Gleiser5}, in the study of Bose-Einstein condensates \cite{Casadio}, and in braneworlds to investigate multi-kink type field configurations \cite{Wilami,MLA}.

The other purpose of this work is to investigate the existence of stationary solutions and calculate the DCC in the $O(3)$-sigma model coupled to the Chern-Simons (CS) field and subject to a kind of nonpolynomial potential. Using the DCC, we investigated the possible existence of phase transitions that can create new topological structures in the model.

We organized this paper as follows: In the Sec. II, a review of the gauged O(3)-sigma model with Chern-Simons term is present. In Sec. III, the definition of the Chern-Simons-Gausson O(3)-sigma model, is implemented. Also, we investigated their stationary solutions using a numerical approach. Then, in Sec. IV, we investigate the DCC and analyze the possible phase transitions of the BPS structures. Finally, in Sec. V, our findings are announced.

%%%%%%%%%%%%%%%%%%%%%%%%%%%%%%%%%%%%%%%%%%%%%%%%%%%%%%%%%%
\section{The gauged $O(3)$-sigma model with Chern-Simons term}
\label{sec:1}

We started our investigation considering a Lagrangian density in flat space-time in $(2+1)$D as proposed in Refs. \cite{ghosh,Belendryasova}, given by
\begin{align}
    \mathcal{L}=\frac{1}{2} D_{\mu}\Phi\cdot D^{\mu}\Phi+\frac{\kappa}{4}\varepsilon^{\mu\nu\lambda}A_{\mu}F_{\nu\lambda}-V(\phi_{3}).
\end{align}
where $V(\phi_{3})$ is the potential and the component $\phi_{3}$ is the field configuration responsible for the spontaneous breaking of the symmetry of the model.

The O(3)-sigma model is well-known in the specialized literature as the result of a mapping of two unit spheres, denoted by $S^{2}$ \cite{Rajaraman}. It is essential to mention that due to this mapping, the field $\Phi$ of the $O(3)$-sigma model must respect the constraint
\begin{align}
\label{sigma}
    \Phi\cdot\Phi=1\rightarrow \phi_{1}^{2}+\phi_{2}^{2}+\phi_{3}^{2}=1.
\end{align}

Due to the choice of a potential of the type $V(\phi_{3})$, the Lagrangian must be invariant to an isorotation of the preference axis $\hat{n}_{3}$, i. e., $\hat{n}_{3} =(0,0,1)$.

We must define the covariant derivative of the model as
\begin{align}
D_{\mu}\Phi=\partial_{\mu}\Phi+A_{\mu}\hat{n}_{3}\times\Phi.
\end{align}

The U(1) nature of the model can be seen by the following identity
\begin{align}
    D_{\mu}\Phi\cdot D^{\mu}\Phi=|(\partial_{\mu}+iA_{\mu})(\phi_{1}+i\phi_{2})|^{2}+\partial_{\mu}\phi_{3}\partial^{\mu}\phi_{3}.
\end{align}
We will assume that our metric signature will be $\eta_{\mu\nu}=$ diag$(+,-,-)$ with $ \varepsilon^{012}=1$; $\mu,\nu=0,1,2$ and $i, j=1,2$.

By investigating the equations of motion, we will have 
\begin{align}
    \textbf{J}^{\mu}=-\Phi\times D^{\mu}\Phi,
\end{align}
where the local current is given by $\textbf{J}^{\mu}=-j^{\mu}\cdot\hat{n}_{3}$ and
\begin{align}
\label{current}
    j^{\mu}=\frac{\kappa}{2}\varepsilon^{\mu\nu\lambda}F_{\nu\lambda}.
\end{align}
The $j^{0}$ component is known as Gauss' law. In this way, the field configurations carry a magnetic flux and a non-zero charge given by $\mathcal{Q}=-\kappa\Phi_{flux}$.

Similarly, the equation of motion for the scalar field will be
\begin{align}
    D_{\mu}D^{\mu}\Phi=-\frac{\partial V}{\partial\Phi}.
\end{align}

Therefore, we can write 
\begin{equation}
    D_{\mu}\textbf{J}^{\mu}=\Phi\times\frac{\partial V}{\partial\Phi}.
\end{equation}

Considering that the $T_{00}$ component of the energy-momentum tensor represents the energy density of the model, we have 
\begin{align}
\label{tensor}
    T_{00}=\mathcal{E}=\frac{1}{2}(D_{1}\Phi)^{2}+\frac{1}{2}(D_{2}\Phi)^{2}+\frac{\kappa^{2}F_{12}^{2}}{2(1-\phi_{3}^{2})}+V,
\end{align}
with the energy that describes the vortex configurations being
\begin{align}
\label{BPSEnergy}
    E=\int\, d^{2}x\, \mathcal{E}.
\end{align}

\section{The Chern-Simons O(3)-sigma model with logarithmic potential}

To study the static field configurations at the BPS bound, i. e., vortex configurations. Let us rewrite the energy density as
\begin{align}
   \label{energy1}
   \mathcal{E}=\frac{1}{2}(D_{i}\Phi\pm\varepsilon_{ij}\Phi\times D_{j}\Phi)^{2}+\frac{\kappa^{2}}{2(1-\phi_{3}^{2})}\bigg[ F_{12}\mp\sqrt{\frac{2(1-\phi_{3}^{2})V}{\kappa^{2}}}\bigg]^{2} \pm4\pi\mathcal{Q}_{0}.
\end{align}

Now, we define the topological charge as
\begin{align}
    \mathcal{Q}_{\mu}=\frac{1}{8\pi}\varepsilon_{\mu\nu\lambda}\bigg[\Phi\cdot D^{\nu}\Phi\times D^{\lambda}\Phi+F^{\nu\lambda}\sqrt{\frac{2\kappa^{2}V}{(1-\phi_{3}^{2})}}\bigg].
\end{align}

In this way, the energy density is
\begin{align}
\label{energy2}
      \mathcal{E}=\frac{1}{2}(D_{i}\Phi\pm\varepsilon_{ij}\Phi\times D_{j}\Phi)^{2}+\frac{\kappa^{2}}{2(1-\phi_{3}^{2})}\bigg[ F_{12}\mp\sqrt{\frac{2(1-\phi_{3}^{2})V}{\kappa^{2}}}\bigg]^{2}\pm4\pi\mathcal{Q}_{0}. 
\end{align}

Considering that the energy is the integration of $T_{00}$ in all space, we will therefore have 
\begin{align}\nonumber
    E=&\frac{1}{2}\int\, d^{2}x\, \bigg\{ (D_{i}\Phi\pm\varepsilon_{ij}\Phi\times D_{j}\Phi)^{2}+\frac{\kappa^{2}}{2(1-\phi_{3}^{2})}\bigg[ F_{12}\mp\sqrt{\frac{2(1-\phi_{3}^{2})V}{\kappa^{2}}}\bigg]^{2}\bigg\}\\
    &\pm 4\pi\int\, d^{2}x\, \mathcal{Q}_{0}.
\end{align}

Allow us to define the BPS energy as
\begin{align}
    E=\pm 4\pi\int\, d^{2}x\, \mathcal{Q}_{0}.
\end{align}

Perceive that the energy of the vortex is limited, i. e.,
\begin{align}
    E\geq E_{BPS}.
\end{align}

At the bound of saturation of the energy, the static field configurations obey the first order Bogomol'nyi equations, given by:
\begin{align}
\label{BPS}
    D_{i}\Phi=\mp\varepsilon_{ij}\Phi\times D_{j}\Phi,
\end{align}
and
\begin{align}
    \label{BPS00}
    F_{12}=\pm\sqrt{\frac{2(1-\phi_{3}^{2})V}{\kappa^2}}.
\end{align}

\begin{figure}[ht!]
\centering
\includegraphics[height=6cm,width=7.5cm]{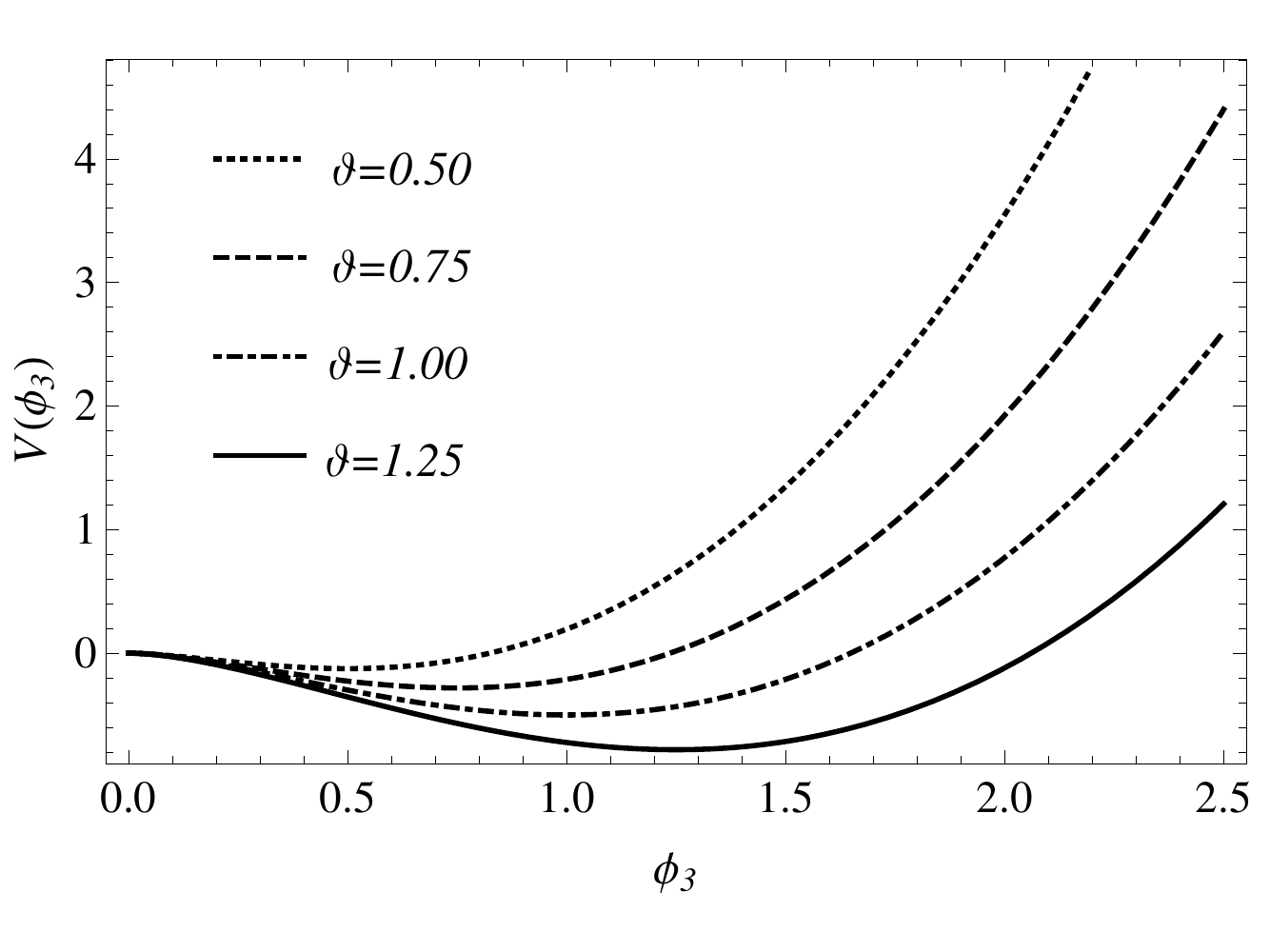}
\vspace{-0.8cm}
\caption{Potential for several values of $\vartheta$.}
\label{fig1}
\end{figure}

At this moment, let us assume that the potential of the theory is given by
\begin{align}
    \label{potential}
    V(\phi_{3})=\frac{\kappa^{2}}{2}\phi_{3}^{2}\text{ln}\bigg(\frac{\phi_{3}^{2}}{\vartheta^{2}}\bigg).
\end{align}
This potential is known as Gausson's potential. In truth, this term is due to the Gaussian form of the soliton-like solutions of the Schr\"{o}dinger equation with logarithmic interaction. Bialynicki-Birula and Mycielski \cite{BBM} were the first to use the term Gausson. Here, the parameters $\kappa$ and $\vartheta$ adjust the dimension of the model, and the factor $\kappa^{2}/2$ is conveniently applied to get the BPS bound. Also, note from Fig. (\ref{fig1}) that the parameter $\vartheta$ is associated with the vacuum state of the model.

The potential (\ref{potential}), allows us to rewrite Eq. (\ref{BPS00}) as
\begin{align}
    \label{BPS1}
    F_{12}=\pm\sqrt{(1-\phi_{3}^{2})\phi_{3}^{2}\text{ln}\bigg(\frac{\phi_{3}^{2}}{\vartheta^{2}}\bigg)}.
\end{align}

\subsection{Static vortex solutions}

Due to the constraint of the $O(3)$-sigma model, we choose a spherical symmetry for the variable field. This means that
\begin{align} 
\label{ansatz1}
\nonumber
    &\phi_{1}=\sin{f(r)} \cos{N\theta},\\
    &\phi_{2}=\sin{f(r)} \sin{N\theta}, \\ \nonumber
    &\phi_{3}=\cos{f(r)}.
\end{align}

Meanwhile, the gauge field is
\begin{align}
\label{ansatz2}
    \textbf{A}(\rho,\theta)=-\frac{Na(r)}{\kappa r}\hat{e}_{\theta}.
\end{align}

Considering the self-dual potential (\ref{potential}), the ansatz for the scalar field (\ref{ansatz1}) and the ansatz for the gauge field (\ref {ansatz2}), we recast Bogomol'nyi's equations as
\begin{align}
\label{BPS3}
    f'(r)=\pm N\frac{a+1}{r}\sin{f(r)},
\end{align}
and
\begin{align}
\label{BPS4}
    a'(r)=\pm\frac{r}{N}\sqrt{\cos{f(r)}^{2}\sin{f(r)}^{2}\text{ln}\bigg(\frac{\cos{f(r)}^{2}}{\vartheta^{2}}\bigg)}.
\end{align}

\subsubsection{Asymptotic analysis}

From now on, we are interested in investigating the topological vortex structures of the model. In other words, we will investigate the solutions of the equations (\ref{BPS3}) and (\ref{BPS4}). To ensure that the variable field has no singularity at the origin, we consider that the field near the origin has the form
\begin{align}
    f(0)=n\pi, \hspace{1cm} n\in\mathbb{N}.
\end{align}
For this behavior of the variable field $f(r)$ near the origin, the gauge field must assume the behavior $a(0)\rightarrow 0$. It is interesting to mention that the solutions of the variable field are symmetric under $f(r)=2\pi$.

We are interested in investigating topological solutions, i. e., solutions with non-zero and finite energy configurations. For this, we are interested in field configurations that respect the conditions
\begin{align}
\label{boundary}
    f(0)=0 \hspace{1cm} \text{and} \hspace{1cm} f(\pi)=\pi.
\end{align}

Considering the condition $f(0)=0$, it is very useful to use the transformation $f(r)=\pi+\xi(r)$. We use the negative sign of the Bogomol'nyi equations and assumed structures with positive vorticity, i. e., $N>0$. Without losing generality, being $\xi(r)\ll 1$, the model assumes solutions of the type
\begin{equation}
    \xi(r)=A_{0}r^{N}.
\end{equation}

Consequently, the solution of the gauge field is
\begin{equation}
    a(r)\simeq-\frac{A_{0}}{N(N+2)(N+3)}r^{N+3}\sqrt{\text{ln}\bigg(\frac{1}{\vartheta}\bigg)}+\mathcal{O}(r),
\end{equation}
where $\mathcal{O}(r)$ are the high order terms of $r$.

On the other hand, if we consider that $f(0)=0$ and $N<0$, we can assume that 
\begin{equation}
    f(r)=\Bar{B_{0}}r^{-N}.
\end{equation}

In this case, the solution for the gauge field is
\begin{equation}
    a(r)\simeq-\frac{\Bar{B_{0}}}{N(2-N)}r^{2-N}\sqrt{\text{ln}\bigg(\frac{1}{\vartheta}\bigg)}+\mathcal{O}(r).
\end{equation}

At infinity, there are two asymptotic behaviors in Bogomol'nyi's equations. First, when $f(\infty)=\pi$, we can again write $f(r)=\pi+\xi(r)$ and obtain the localized energy solutions with $a(\infty)=\eta_{1}$. That way, we can assume
\begin{equation}
    \xi(r)=C_{\infty}r^{N(1-\eta_{1})}.
\end{equation}

As result, we have that
\begin{equation}
    a(r)\simeq-\frac{C_{\infty}}{[N(1-\eta_{1})+2]}r^{[N(1-\eta_{1})+2]}\sqrt{\text{ln}\bigg(\frac{1}{\vartheta^{2}}\bigg)}-\eta_{1}.
\end{equation}

Finally, when analyzing the conditions:
\begin{align}
    f(\infty)=0 \hspace{1cm} \text{and}
    \hspace{1cm} a(\infty)=\eta_{2}
\end{align}
with $N<0$, we have
\begin{equation}
    f(r)=\Bar{C}_{\infty}r^{N(1+\eta_{2})}.
\end{equation}

Consequently, we obtain 
\begin{equation}
    a(r)=-\frac{\Bar{C}_{\infty}}{N[N(1+\eta_{2})+2]}r^{[N(1+\eta_{2})+2]}\sqrt{\text{ln}\bigg(
    \frac{1}{\vartheta^{2}}\bigg)}+\eta_{2}.
\end{equation}
The parameters $\eta_{1}$ and $\eta_{2}$ indicate whether the solutions are topological or non-topological.

\subsubsection{The numerical results}

\begin{figure}[ht!]
\centering
\includegraphics[height=6cm,width=7.5cm]{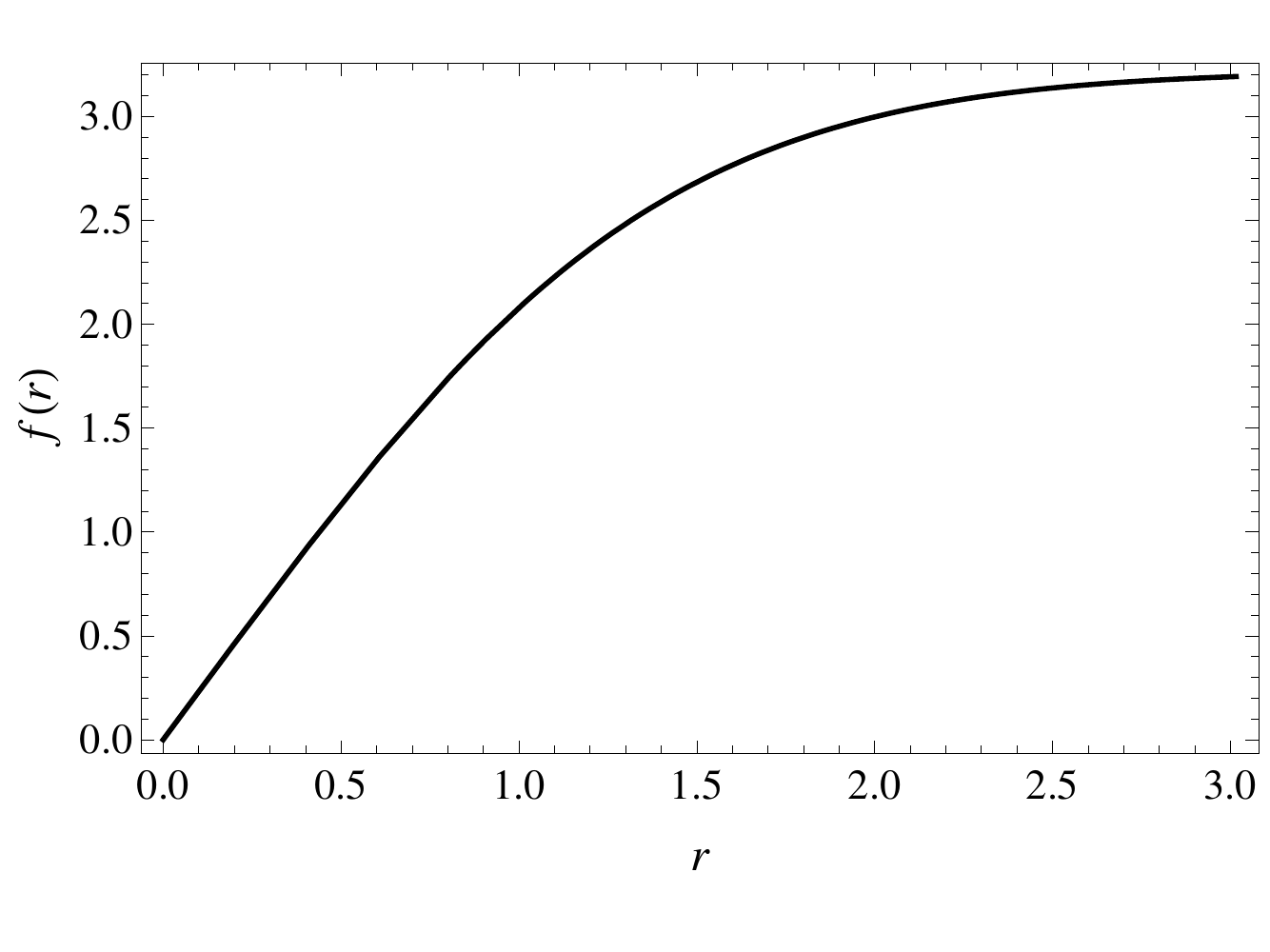}
\includegraphics[height=6cm,width=7.5cm]{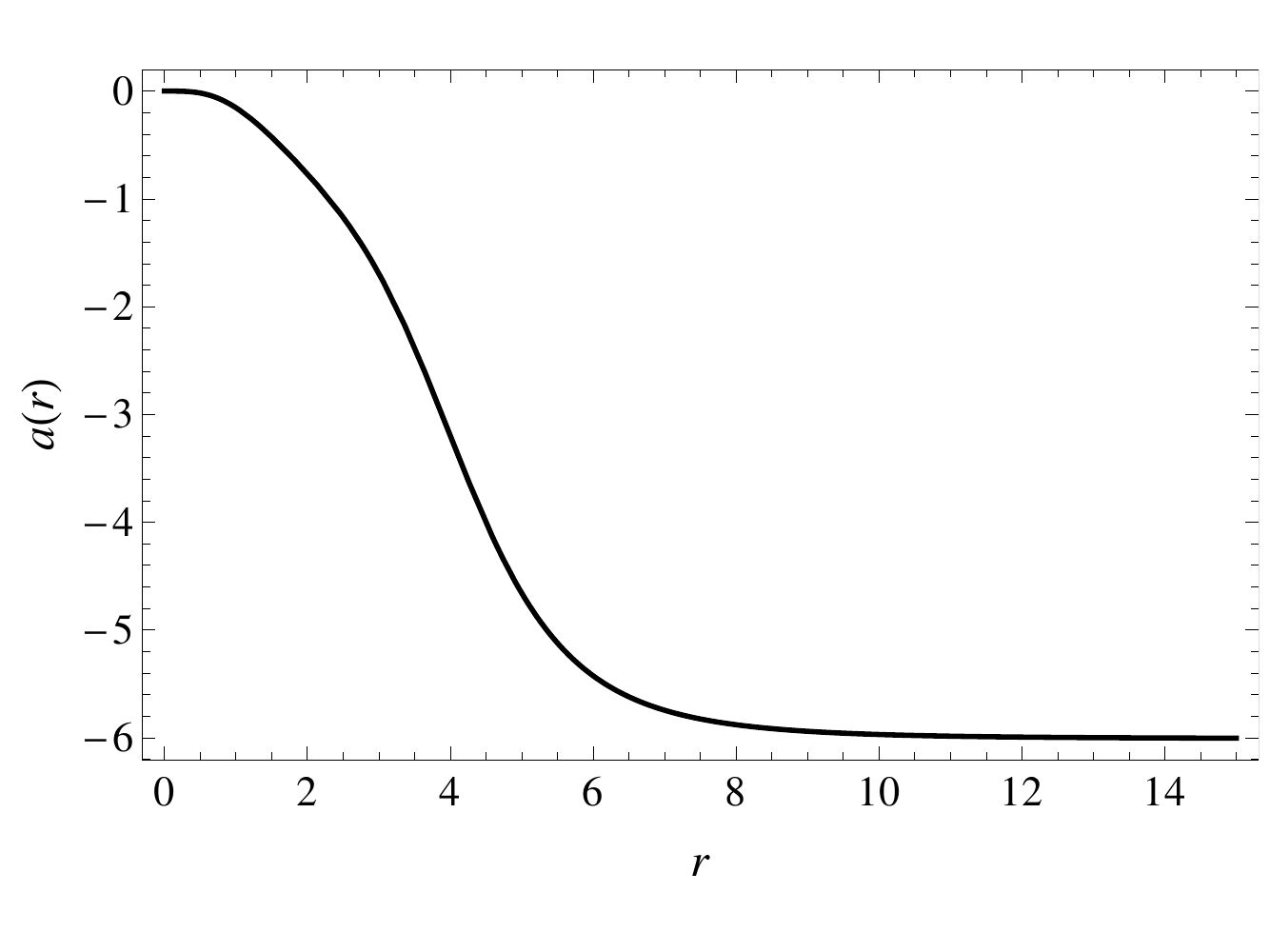}\\
\vspace{-0.75cm}
\begin{center}
\hspace{0.5cm} (a) \hspace{7.5cm} (b)
\end{center}
\vspace{-0.5cm}
\caption{Topological solution of the $O(3)$-sigma model with logarithmic potential. (a) The result of the scalar field. (b) The result of the gauge field.}
\label{fig2}
\end{figure}

We return our attention to the numerical study of the scalar and gauge fields. To achieve our purpose, we consider the fields described by respective Eqs. (\ref{BPS3}) and (\ref{BPS4}). Also, we will use a numerical interpolation method with the topological conditions (\ref{boundary}). It is shown in Fig. \ref{fig2}(a) the result of the scalar field.

On the other hand, with the help of the equation (\ref{BPS4}), we can obtain the graphic behavior of the Chern-Simons field associated with vortex structures and represented in Fig. \ref{fig2}(b).

\begin{figure}[ht!]
    \centering
    \includegraphics[height=6cm,width=7.5cm]{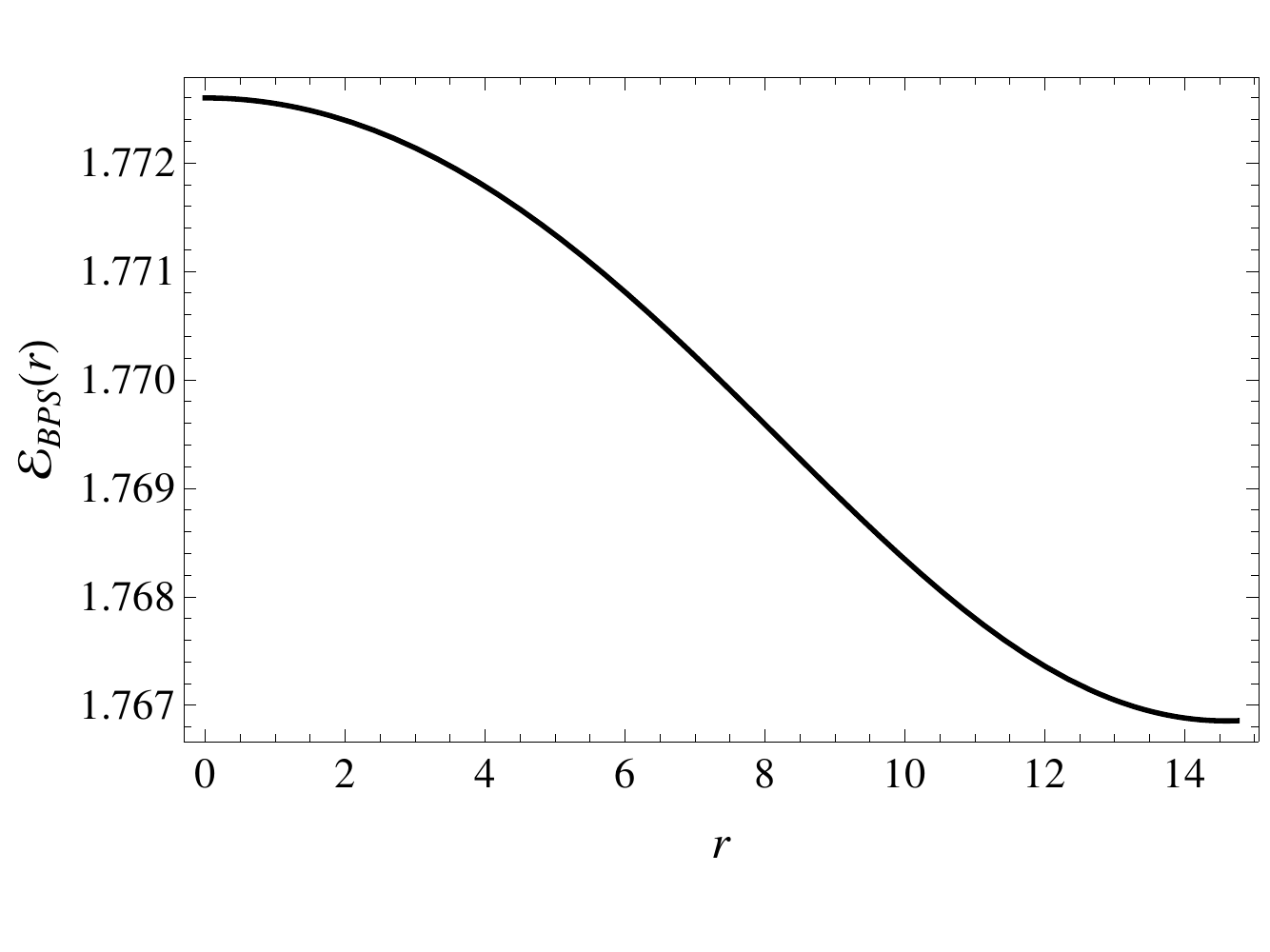}
    \includegraphics[height=6cm,width=7.5cm]{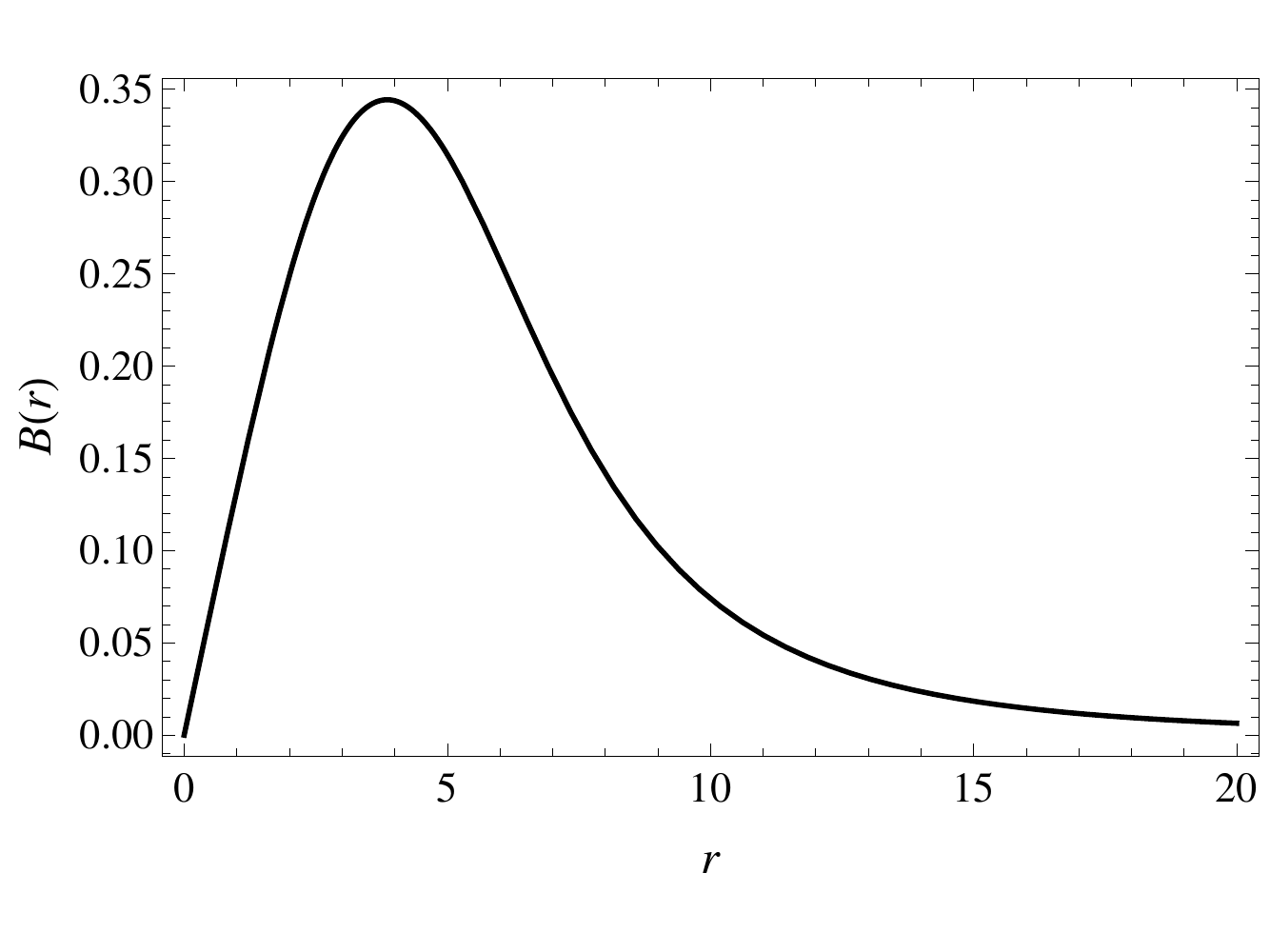}\\
    \vspace{-0.75cm}
    \begin{center}
    \hspace{0.5cm} (a) \hspace{7.5cm} (b)
    \end{center}
    \vspace{-0.5cm}
    \caption{(a) The BPS energy density of the vortex. (b) The magnetic field of the vortex.}
    \label{fig3}
\end{figure}

Using the numerical solutions of the variable field and the gauge field, we can describe the physical properties of the structure, namely: BPS energy density and magnetic field. Considering the solutions displayed in Figs. \ref{fig2}(a) and \ref{fig2}(b) we obtain the BPS energy density of the vortex. Let us remember that to obtain the BPS energy density is necessary to use the equation  (\ref{tensor}). Fig. \ref{fig3}(a), is exposed the numerical result of the BPS energy density.

Adopting the expression (\ref{BPS1}), we obtain the magnetic field (Fig. \ref{fig3}(b)) of the vortex structures. The numerical results allow reaching the kink-like topological vortex structures. We can see that this structure has a finite and positive-defined energy density with very significant intensity. This result is because the vortices of the model interact with the Chern-Simons field generating a flux due to the magnetic induction shown in Fig. \ref{fig3}(b).

Finally, to calculate DCE and its variants is usual to use the BPS vortex solutions, see Ref. \cite{LA12}. It is essential to highlight that this is because these quantities, in truth, are defined by a Fourier transform of the energy density (in our case, the BPS energy density). Indeed, the results of the CE and its variants provide information about the model parameters so that it is possible to detect stable field configurations \cite{Gleiser,Gleiser1,Gleiser2}, see Sec. IV. In the study of BPS structures, the CE will inform us which values of the BPS topological configurations are stable or more likely. The information about the DCC and phase transitions will be relevant to complement our discussion of the BPS solutions obtained in this section. In general, the study of the information theory in this context will give us information about the stability of the solutions of the BPS vortex found.

\section{DCC of the vortex}

The configurational entropy (CE) gives us a measure of the informational complexity of a localized field configuration. The CE, DCE, and DCC are amounts formulated as Fourier transforms of the energy density, see Refs. \cite{Gleiser,Gleiser1,Gleiser2,Wilami}. In fact, the CE has been studied intensively in several cosmological scenarios \cite{correia,Correia5,CLee}. With this in mind, to understand the vortex structure and the possible existence of phase transitions of the  BPS solutions, we propose the study of the DCC of the model. 

From Refs. \cite{Gleiser,Gleiser1,Gleiser2,Gleiser3}, we define the concept of differential entropy as
\begin{align}
    \mathcal{S}=\int\, \rho(\mat{\omega})\ln[\rho(\mat{\omega})]\,d\mat{\omega}.
\end{align}
It is worth noting that the DCC is not invariant under a change of coordinates. As a consequence of this, the probability density transforms like a scalar density under the coordinate transformations $x\to\tilde{x}$, see Ref. \cite{GS1}.

Considering the localized energy density $\mathcal{E}(\textbf{r})$, we define in D-dimensional space the decomposition of the wave modes as
\begin{align}
    \mathcal{F}(\mat{\omega})=(2\pi)^{-(d/2)}\int \mathcal{E}(\textbf{r})\text{e}^{-i \mat{\omega}\cdot\textbf{r}}\, d^d \textbf{r}.
\end{align}

The contribution of wave mode to a $\vert\mat{\omega}_\star\vert$ is the modal fraction, namely,
\begin{align}
    f(\mat{\omega})=\frac{\vert\tilde{\mathcal{F}}(\mat{\omega})\vert^2}{\vert\tilde{\mathcal{F}}(\mat{\omega}_\star)\vert^2}.
\end{align}

The DCC is defined as
\begin{align}
    \mathcal{S_C}=-\int\, f(\mat{\omega})\ln[f(\mat{\omega})]\, d^d\mat{\omega},
\end{align}
note that the above integral is positive definite since $g(\mat{\omega})\leq 1$.

Due to the symmetry of our model, it is convenient to write the DCC in spherical symmetry, which lets us remember that the hyperspherical Fourier transform is given by
\begin{align}\label{defF}
    \mathcal{F}=\omega^{1-\frac{d}{2}}\int_{0}^{\infty}\, \mathcal{E}(r)J_{\frac{d}{2}-1}(\omega r)r^{d/2}\, dr,
\end{align}
with $\mathcal{E}(r)$ the BPS energy density of the vortex and $J_\nu$ the Bessel function.

Using (\ref{defF}), we have that the DCC is
\begin{align}
    \mathcal{S_C}=\frac{2\pi^{d/2}}{\Gamma(\frac{d}{2})}\int_{0}^{\infty}\, \omega^{d-1} f(\omega)\ln[f(\omega)]\, d\omega,
\end{align}
where the entropic density $\rho(\omega)$ is the integrand of the DCC.

In the context of the braneworlds, the above quantity can describe new topological structures, e. g. multi-kink solutions, which are the fruits of multiples phase transition \cite{Wilami}. Using the DCC to study the topological structures, we want to understand whether our vortex solutions are unique and whether it permits multiple transitions.

The calculation of the DCC is not easy since our energy density of the O(3)-sigma is given by 
\begin{align}
    \mathcal{E}=\frac{1}{2}D_i\Phi\cdot D^{i}\Phi+\frac{\kappa^2 F_{12}^{2}}{2(1-\phi_3)}+\frac{\kappa^2}{2}\phi_3\text{ln}\frac{\phi_{3}^{2}}{\vartheta^2},
\end{align}
and
\begin{align}
    \Phi=\phi_{i}(f(r);r)\hat{\text{e}}\,^{i} \, \, \, \, \, \text{with} \, \, \, \, \, i=1,2,3.
\end{align}

\begin{figure}[ht!]
    \centering
    \includegraphics[height=6cm,width=7.5cm]{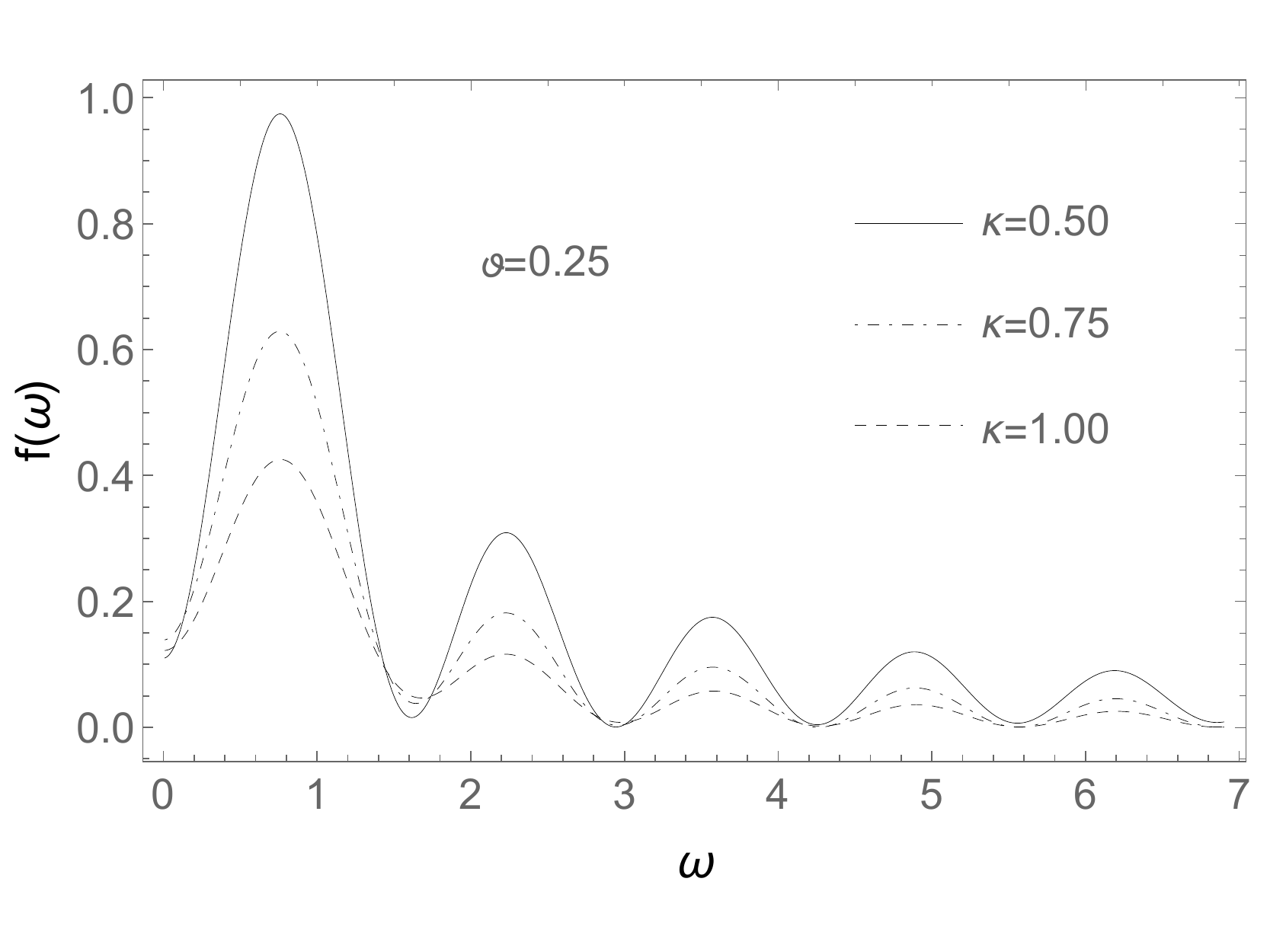}
    \includegraphics[height=6cm,width=7.5cm]{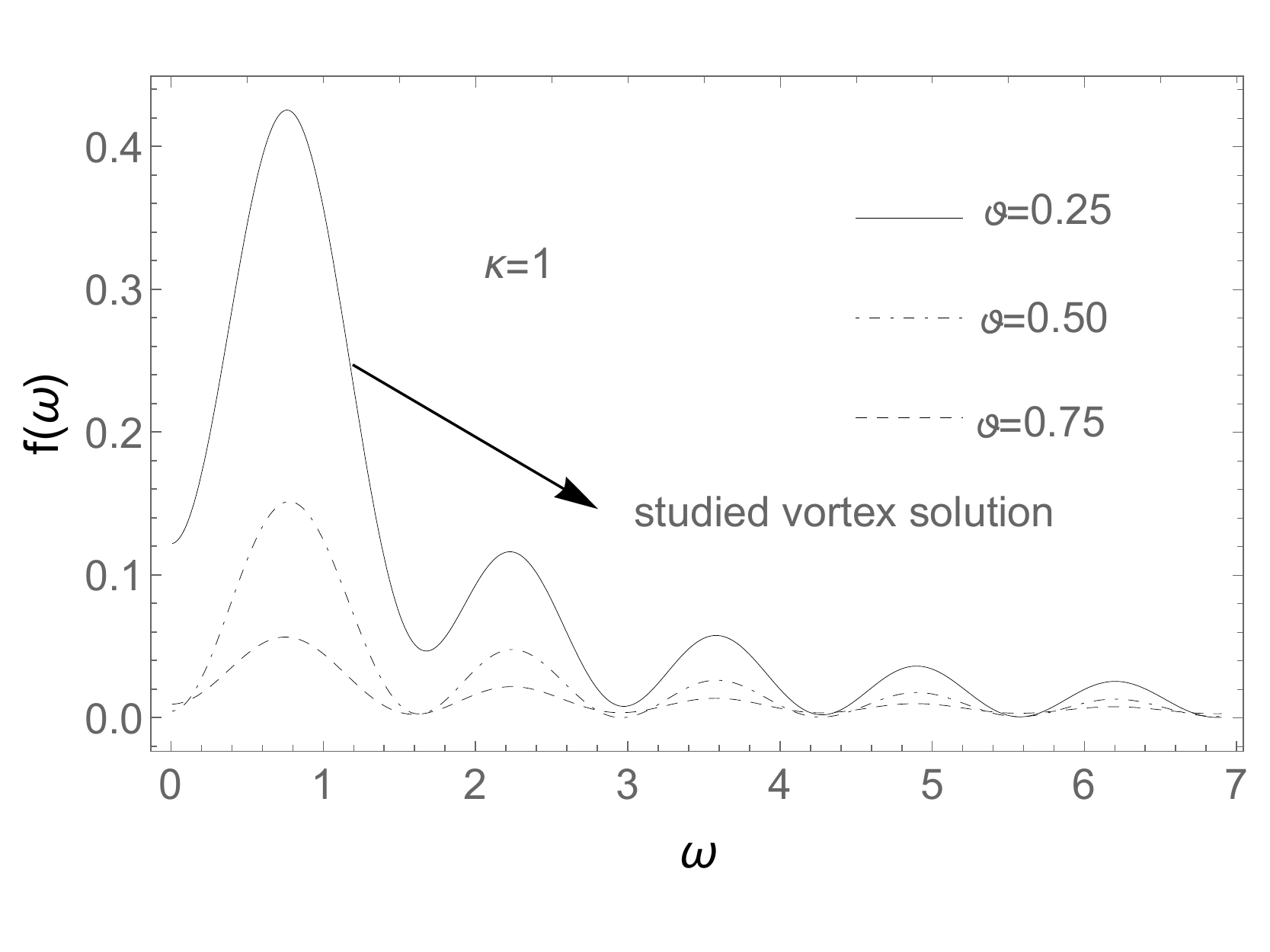}\\
    \vspace{-0.75cm}
    \begin{center}
    \hspace{0.5cm} (a) \hspace{7.5cm} (b)
    \end{center}
    \vspace{-0.5cm}
    \caption{(a) Modal fraction
    when $\vartheta$ is constant. (b) Modal fraction when $\kappa$ is constant.}
    \label{fig6}
\end{figure}

Although of the difficulties, we evaluate the modal fraction numerically of the model and show the results in Figs. \ref{fig6}(a) and (b). To obtain the modal fraction, we consider the numerical solutions of Eqs. (\ref{BPS3}) and (\ref{BPS4}). First, we perceive that the parameter $\vartheta$, which relates the potential to the magnetic field, will directly influence the modal fraction. In this way, the increase of the parameter $\vartheta$ produces an exponential decrease of the modal fraction (amplitude) of the BPS vortex. By simulation, we also noticed the amplitude of the first node in figs. \ref{fig6}(a) and (b) correspond to the solution discussed in the previous section.

\begin{figure}[ht!]
    \centering
    \includegraphics[height=6cm,width=7.5cm]{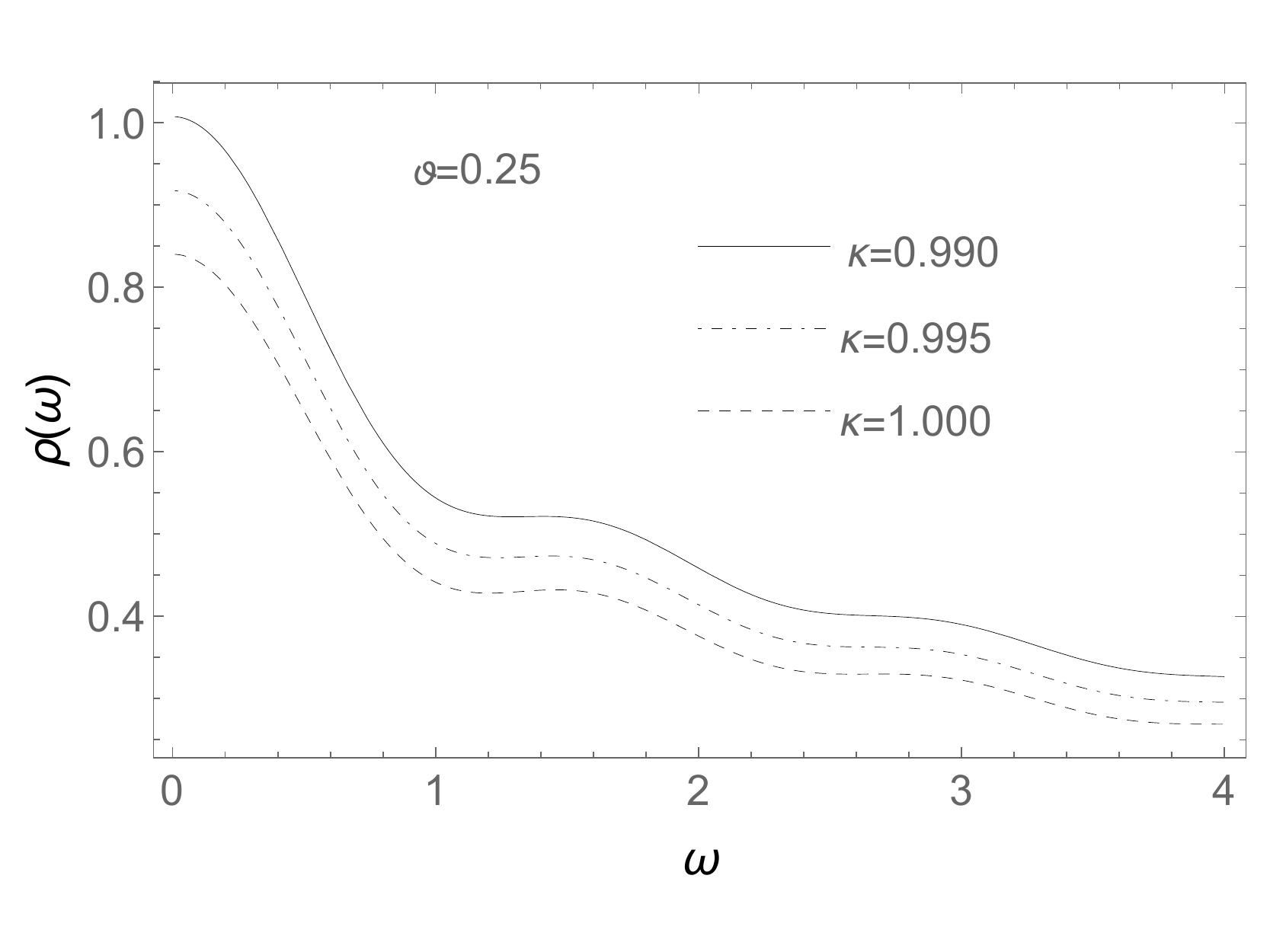}
    \includegraphics[height=6cm,width=7.5cm]{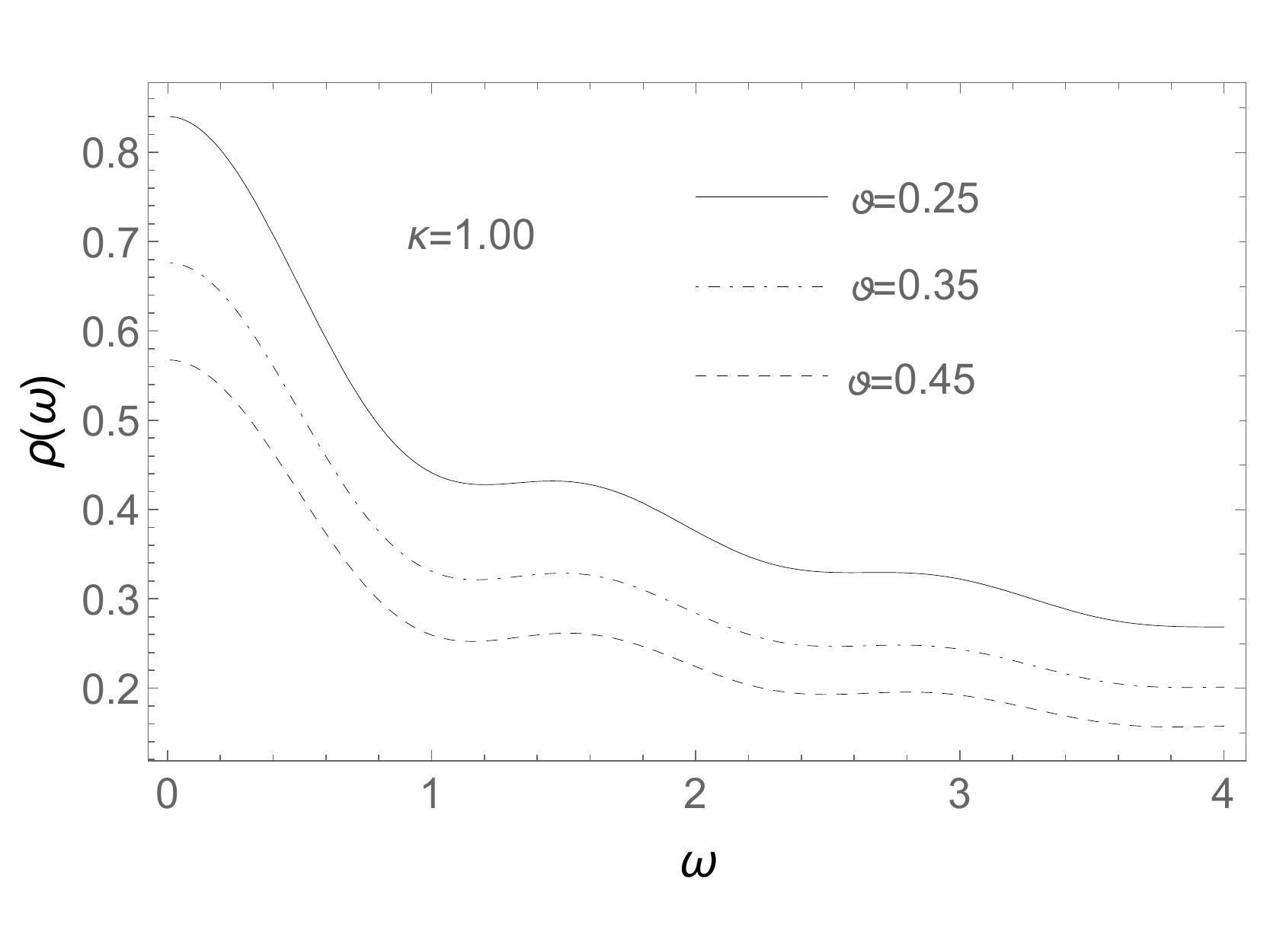}\\
    \vspace{-0.75cm}
    \begin{center}
    \hspace{0.5cm} (a) \hspace{7.5cm} (b)
    \end{center}
    \vspace{-0.5cm}
    \caption{(a) Configurational entropy density when $\vartheta$ is constant. (b) Configurational entropy density  when $\kappa$ is constant.}
    \label{fig7}
\end{figure}

Using the numerical solution of the modal fraction, we investigate the configurational entropy density (entropic density) of the model. In the neighborhood of the core of the BPS vortex ($r\approx 0 $), we observed entropic densities are higher for parameters $\kappa$ and $\vartheta$ fixed. In fact, due to the localized structure being in $r=0$, the entropic density is high in this region. The parameter $\vartheta$ is responsible for the control entropic density and the magnetic field. In truth, when this parameter increases, the magnetic field decreases, leading to decay of the modal fraction (Figs. \ref{fig6}) and the entropic density (Figs. \ref{fig7}).

\begin{figure}[ht!]
    \centering
    \includegraphics[height=6cm,width=7.5cm]{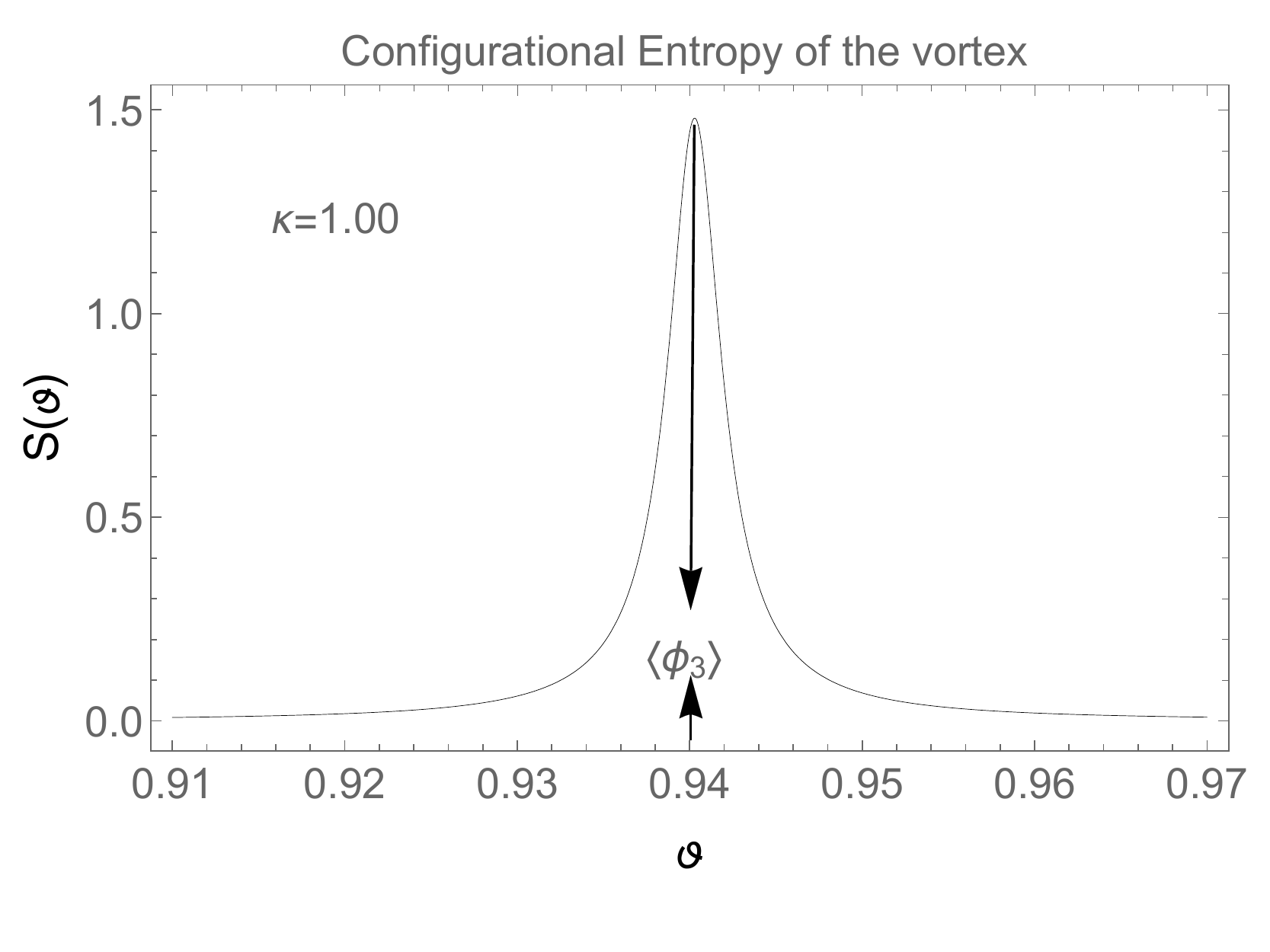}
    \vspace{-0.8cm}
    \caption{Configurational entropy of the BPS vortex.}
    \label{fig8}
\end{figure}

Finally, the results of the DCC of the BPS vortex (Fig. \ref{fig8}) allow us to observe that the Chern-Simons-Gausson O(3)-sigma model admits only a type of solution, namely, the kink-like solution. In Fig. \ref{fig2}(a), we exposed this solution. Due to the DCC profile of the O(3)-sigma model, we observed that the Chern-Simons-Gausson O(3)-Sigma model does not support multiple phase transitions. Also, it is possible to see that the critical point of the DCC corresponds to the vacuum expected value of the theory.

\section{Concluding remarks}

In this work, we studied the vortex topological structures of the O(3)-sigma model. In our model, we coupled the field configuration of the sigma model to the Chern-Simons with logarithmic interaction. We show that the topological structures are kink-like solutions due to the spherical symmetry of the problem. As a consequence of this result, the vortices generated are magnetically charged and have a magnetic flux given by $\Phi_{flux}=-\mathcal{Q}/\kappa$. Finally, numerically was observed that the vortices generated are quite energetic. Indeed, this is due to the contribution of the logarithmic potential and the Chern-Simons field.

Once we obtain the BPS solution of the sigma model, we find the spatial profile of the energy density of the model. We investigated whether the studied vortices admitted multiple phase transitions. We observed that the DCC assumed a profile type of a directional delta-function centered approximately on the expected value of the vacuum state. Due to this profile of the DCC, it is clear that the model does not support multiple phase transitions. In fact, due to the behavior of the DCC and the probability densities, we note that the model supports other solutions, however, these solutions will differ only ``how quickly'' the fields evolve into a vacuum state.

By numerical analysis, we perceived entropy densities are higher where the vortex is localized. Indeed, examining the parameters $\kappa$ and $\vartheta$ note that the influence of the vortex structures makes their energetic flux higher (or smaller) as the field is more (or less) localized.

\section*{Acknowledgment}

The authors thank the Conselho Nacional de Desenvolvimento Cient\'{i}fico e Tecnol\'{o}gico (CNPq), grant No. 308638/2015-8 (CASA), and Coordena\c{c}\~{a}o de Aperfei\c{c}oamento do Pessoal de N\'{i}vel Superior (CAPES) for financial support.
%%%%%%%%%%%%%%%%%%%%%%%%%%%%%%%%%%%%%%%%%%%%%%%%%%%%%%%%%%

\end{document}